\documentclass[orivec]{llncs} 

\usepackage{times}
\usepackage{mathptmx}       
\usepackage{helvet}         
\usepackage{courier}        
\usepackage{type1cm}        
\usepackage{makeidx}         
\usepackage{graphicx}        
\usepackage{multicol}        
\usepackage[bottom]{footmisc}

\usepackage{amsmath}
\usepackage{amssymb}

\usepackage{hhline}
\usepackage{tabularx}
			
\usepackage{url}

\usepackage{epstopdf}

\usepackage{xspace}

\usepackage{xcolor}

\newcommand{\nop}[1]{}

\newcommand{\relnames}{\textit{Rel}\xspace }
\newcommand{\attnames}{\textit{Att}\xspace }
\newcommand{\dom}{\textit{Dom}\xspace }
\newcommand{\varnames}{\textit{Var}\xspace}
\newcommand{\D}{\mathcal{DS}}
\def\R{{{\mathbb{R}}}}
\def\I{{\mathcal{I}}}

\newcommand{\IC}{\mathcal{C}}

\newcommand{\DC}{\mathcal{C}}
\newcommand{\MI}{{\sf{MI}}}
\newcommand{\MC}{{\sf{MC}}}
\newcommand{\Prob}{{\sf{Problematic}}}
\newcommand{\Free}{{\sf{Free}}}
\newcommand{\Safe}{{\sf{Safe}}}
\newcommand{\Self}{{\sf{Selfcontradictions}}}
\newcommand{\Contradictory}{{\sf{Contradictory}}}
\newcommand{\Conflictbase}{{\sf{Conflictbase}}}

\newcommand{\LV}{\textbf{LV}}
\newcommand{\UV}{\textbf{UV}}
\newcommand{\EV}{\textbf{EV}}
\newcommand{\IM}{\textbf{IM}}

\usepackage{pifont}
\newcommand{\cmark}{\ding{51}}%
\newcommand{\xmark}{\ding{55}}%

\begin{document}
\title{Inconsistency Measures for Relational Databases}

\author{Francesco Parisi$^1$ John Grant$^2$,}
\institute{$^1$ DIMES Department University of Calabria, Italy, \\
email:  fparisi@dimes.unical.it\\
$^2$ University of Maryland at College Park, USA, \\
email: grant@cs.umd.edu}

\maketitle

\begin{abstract}
In this paper, building on work done on measuring inconsistency in knowledge bases, 
we introduce inconsistency measures for databases.
In particular, focusing on databases with denial constraints,
we first consider the natural approach of virtually transforming 
a database into a propositional knowledge base and then applying well-known measures.
However, using this method, tuples and constraints are equally considered 
in charge of inconsistencies.
Then, we introduce a version of inconsistency measures blaming database tuples only, i.e., treating integrity constraints as irrefutable statements.

We analyze the compliance of database inconsistency measures 
with standard rationality postulates and find interesting 
relationships between measures.
Finally, we investigate the complexity of the inconsistency measurement 
problem as well as of the problems of
deciding whether the inconsistency is lower than, greater than, 
or equal to a given threshold.
\end{abstract}

\section{Introduction}

There is a growing number of applications where inconsistent information arises, often because data are obtained from multiple sources~\cite{CalauttiCFFFGMPPSTZ18}.
This has led to an extensive body of work on handling inconsistent 
data, and in particular inconsistent databases.
Important approaches for dealing with inconsistent databases include for instance consistent query answering frameworks~\cite{ArenasBC99,CalauttiLP18}, data repairing~\cite{AfratiK09,KimelfeldLP17,LivshitsK17,MartinezPPSS14},
as well as interactive data repairing and cleaning systems (e.g.~\cite{FazzingaFFP06,Hao0LHTF17,HeVSLMPT16}).

However, very little work has been done on measuring inconsistency
in databases, a problem which on the other hand has been extensively investigated for knowledge bases (KBs).
Measuring the amount of inconsistency in a database, or more in general in a knowledge base, can help
in understanding the primary sources of such conflicts as well as
devising ways to deal with them.
Furthermore, it makes it possible to compare the amount of
inconsistency between various chunks of information.
Although the idea of measuring inconsistency was introduced more
than 40 years ago, in \cite{Grant78}, at that time it did not seem to
be an important issue.
The problem became more noticeable in the 1990s when it became possible 
to store large amounts of information.
It was only in the early 2000s when several AI researchers started to
investigate this issue systematically~\cite{Knight02}.
The bulk of this work since then has been for propositional knowledge
bases, that is, where the information was presented as a set of
formulas in propositional logic.
In the last couple of years the work has been extended to other
frameworks.
The book, \cite{GrantMartinez}, surveys what has been done so far and
gives some extensions.

\vspace*{1mm}
\noindent
\textbf{Contribution.}
In this paper, we introduce inconsistency measures for relational databases with denial constraints.
In particular, 
starting with the work that has been done over the past
nearly 20 years on measuring inconsistency for propositional knowledge
bases, we make the following contributions.
\begin{itemize}
\item
We first extend propositional inconsistency measures to databases via a transformation from a given database with denial constraints to a propositional knowledge base that preserves inconsistency, virtually allowing the application of any propositional inconsistency measure to a database.
We call such measures 
\textit{propositional inconsistency measures for databases}, and 
denote them as $I_{x}$ with $x\in\{B,M,\#,P,A,H,{nc},{hs},{C},{\eta}\}$
(see Definition~\ref{def:incmeas}).
\item
However, interpreting a database as a knowledge base means treating integrity constraints as propositional formulas having the same importance of
tuples, which is in turn reflected in the way inconsistency is measured. Thus,
for each propositional measure $I_x$,
we introduce a \textit{database inconsistency measure} $\I_x$
that measures inconsistency by blaming database tuples only (Section~\ref{sec:second-method}).
\item
For both versions of the inconsistency measures, $I_x$ and $\I_x$, 
we check for compliance of well-known rationality postulates, showing which postulates that are unsatisfied in the propositional case become satisfied in the database setting, and which ones remain unsatisfied. Interestingly, in some cases, compliance comes from the fact that some measures become identical in our setting (i.e., $I_{hs}$ collapses to $I_B$, 
and $\I_C$ collapses to $\I_H$).
Tables~\ref{tab:postulates1} and~\ref{tab:postulates2} summarize the results obtained.
\item
Finally, we investigate the data complexity of the problems of deciding whether a given value is lower than (\LV), upper than (\UV), or equal to (\EV) the inconsistency measured  for a given database using a 
given inconsistency measure $\I_x$.
A summary of the results obtained for these problems, as well as for the problem of computing the actual value of an inconsistency measure (\IM\ problem), is reported in Table~\ref{tab:summaryRepair}.
Interestingly, while measures $\I_B$, $\I_M$, $\I_\#$, $\I_P$ become tractable in the database setting, and the complexity of $\I_A$ decreases, measures $\I_H$ and $\I_\eta$ remain hard as in the propositional case~\cite{ThimmWallnerKR16}
even under data complexity.
\end{itemize}

\section{Preliminaries}\label{sec:preliminaries}

We first briefly review inconsistency measures for propositional knowledge bases.
For our review we will rely on the survey presented in~\cite{thimm18}
which lists most of the proposed inconsistency measures, rationality
postulates, and their satisfaction for propositional logic.
In Section~\ref{sec:review} we present the definitions we will need later, and refer the
reader for details and a complete picture of the situation
to that survey and references inside.
After this, we give the notation we use for databases in Section~\ref{sec:notationDBs}. 

\subsection{Review of Inconsistency Measures for Knowledge Bases}\label{sec:review}

The idea of an inconsistency measure is to assign a number to a knowledge base that measures its inconsistency.
Actually, there are two main types of inconsistency measures:
an \textit{absolute} measure measures the total amount of inconsistency; a
\textit{relative} measure is a ratio of the amount of inconsistency with
respect to the size of the knowledge base.
In the literature there is sometimes confusion between these two
types: in this paper we will be dealing with absolute measures as
these have been studied in more detail.

We start with a propositional language of formulas composed from a
countable set of atoms, the fundamental propositions, and the 
connectives $\wedge$, $\vee$, and $\neg$.
We write ${\cal K}$ for the set of all knowledge bases (KBs), i.e.
the set of all finite sets of formulas in the language.
We write $K$ for an individual KB.
$2^{X}$ is the set of all subsets (the power set)
of any set $X$.
An inconsistency measure 
gives each KB a nonnegative real number or infinity.

For a knowledge base $K$,
$\MI(K)$ is the set of minimal inconsistent subsets of $K$,
and $\MC(K)$ is the set of maximal consistent subsets of $K$.
Also, if $\MI(K) = \{ M_1, ..., M_n \}$ 
then $\Prob (K)$ = $M_1 \cup ... \cup M_n$,
and ${\Free}(K)$ = $K \setminus \Prob(K)$.
A free formula 
is not involved in an essential way in any inconsistency,
while a problematic formula is so involved in at least one inconsistency.
Furthermore, a free formula is called {\em safe} if its atoms differ from the
atoms of all the other formulas (in $K$);
we use $\Safe(K)$ to denote the set of safe formulas.
A formula in $K$ that is individually inconsistent (e.g. $p \wedge \neg p$) is called a selfcontradiction. 
We use $\Self(K)$ to denote the set of selfcontradictions of $K$.

Now we are ready to define the inconsistency measure concept.

\begin{definition}[Inconsistency Measure]
\label{def:inconsistencymeasure}
A function $I: {\cal K} \rightarrow \mathbb{R}^{\geq 0}_\infty$ is an 
{\bf inconsistency measure} 
if the following two conditions hold for all $K,K'\in\mathcal{K}$:
\begin{description}
\item[Consistency] $I(K) = 0$ iff $K$ is consistent.
\item[Monotony] If $K\subseteq K'$, then $I(K)\leq I(K')$.
\end{description}
\end{definition}

Consistency and Monotony are called (rationality) postulates.
Postulates are desirable properties for inconsistency measures
and we will present additional ones later.
However, we require that a function on KBs must at least satisfy
these two postulates in order to be called an inconsistency measure.
Consistency means that all and only consistent KBs get measure $0$.
Monotony means that the enlargement of a KB cannot decrease its measure.
Monotony is not appropriate for relative measures where the ratio of
inconsistency may decrease with the addition of consistent information;
however, it is appropriate for absolute measures.

For some of the inconsistency measures that we will present in Definition~\ref{def:incmeas}, we need additional definitions 
that we present next.

We will be dealing both with classical (two-valued) interpretations
for the atoms as well as three-valued (3VL) interpretations.
A {\em classical interpretation} assigns each atom (in the KB) 
the value T (\textit{true}) or F (\textit{false}).
Using the usual definitions of logical connectives each formula is
also assigned a truth value.
For a \textbf{3VL interpretation} each atom gets one of the three values
T (\textit{true}), F (\textit{false}), or B (\textit{both}).
The logical connectives are extended to 3VL interpretations as shown in
Table~\ref{tab:3VL}, using Priest's three valued logic.
In the classical case, an interpretation is a {\em model} for a set of
formulas if no formula gets the value F.
The same condition is used for 3VL, but now, in addition to T the value B is
also allowed.
We use ${\sf Models}(K)$ to denote the set of 3VL models for a knowledge base $K$.
Also, for a 3VL interpretation $i$ we define 
${\Conflictbase}(i) = \{ a \mid i(a) = B \}$,
the atoms that have truth value B.
In that sense a classical interpretation is a special case of a 3VL
interpretation for which ${\Conflictbase}(i) = \emptyset$.

A \textbf{hitting set}  $H$ \textbf{for the interpretations} of a KB $K$ is a subset of 
the set of the classical interpretations for $K$ such that 
for every $\phi \in K$ there is an interpretation 
$i \in H$ such that $i(\phi) = T$.

\begin{table}
\begin{center}
\begin{tabular}{|c|c|c|c|c|c|c|c|c|c|}
\hline
\multicolumn{1}{|c|}{Formula} & \multicolumn{9}{c|}{Truth value} \\
\hline
$\phi$ & $T$ & $T$ & $T$ & $B$ & $B$ & $B$ & $F$ & $F$ & $F$ \\
$\psi$ & $T$ & $B$ & $F$ & $T$ & $B$ & $F$ & $T$ & $B$ & $F$ \\
$\phi \vee \psi$ & $T$ & $T$ & $T$ & $T$ & $B$ & $B$ & $T$ & $B$ & $F$ \\
$\phi \wedge \psi$ & $T$ & $B$ & $F$ & $B$ & $B$ & $F$ & $F$ & $F$ & $F$ \\
$\neg \phi$ & $F$ & $F$ & $F$ & $B$ & $B$ & $B$ & $T$ & $T$ & $T$ \\
\hline
\end{tabular}
\vspace*{2mm}
\caption{Truth table for three valued logic (3VL). 
This semantics extends the classical semantics with a third truth value, $B$, 
denoting ``inconsistency''.
Truth values on columns 1, 3, 7, and 9, give the classical semantics, 
and the other columns give the extended semantics.
\label{tab:3VL}}
\end{center}
\end{table}

A \textbf{PSAT (probabilistic satisfiability) instance} is a set,
$\Gamma=\{P(\phi_i)\geq p_i\mid 1\leq i\leq m\}$, 
that assigns probability lower bounds to a set $\{ \phi_1, \ldots, \phi_m\}$
of formulas; therefore $0 \leq p_i \leq 1$ for $1 \leq i \leq m$.
A \emph{probability function} over a set $X$ is a function $\pi:X\to [0,1]$ such that $\sum_{x \in X}\pi(x)=1$. 
Let $Int$ be the set of all classical interpretations 
(over $K$) and $\pi$ a probability function over $Int$.
The probability of a formula $\phi$ according to $\pi$ 
is the sum of the probabilities assigned to the interpretations assigning $T$ to $\phi$, that is,
$P_\pi(\phi)=\sum_{i \in Int, i(\phi)=T} \pi(i) $ 
for every formula $\phi \in K$.
A PSAT instance is {\em satisfiable} if there is a probability function
$\pi$ over $Int$ such that $P_\pi(\phi_i) \geq p_i$ for all $1 \leq i \leq m$.

Now we are ready to define the propositional inconsistency measures
we will consider in this paper.
Below the definition we briefly explain the meanings of these measures.

\begin{definition}[Propositional Inconsistency Measures]
\label{def:incmeas}
For a knowledge base $K$, the inconsistency measures 
$I_B$, $I_M$, $I_{\#}$, $I_P$, $I_A$, $I_H$, $I_{nc}$, $I_{hs}$, $I_C$, and $I_\eta$ are such that
\begin{itemize}
\item $I_B(K) = 1$ if $K$ is inconsistent and $I_B(K) = 0$ 
if $K$ is consistent.

\item $I_M(K) = | {\sf MI}(K) |$.

\item $I_{\#}(K) = \left\{ 	\begin{array}{ll} 
					0 & \mbox{if } $K$ \mbox{ is } \mbox{consistent}, \\ 
					\sum_{X \in {\sf MI}(K)} \frac{ 1 }{ | X | } & \mbox{otherwise}. 
					\end{array} \right. $
					
\item $I_P(K) = | {\sf Problematic}(K)  |$.

\item $I_A(K) = (| {\sf MC}(K) | + | \Self(K) |) -1$.

\item $I_H(K) = {\sf min} \{ |X| \mid X \subseteq K \mbox{ and } \forall M \in {\sf MI}(K) (X \cap M \neq \varnothing) \}$.

\item $I_{nc}(K)= |K|-\max\{n\mid \forall K'\subseteq K: 
|K'|=n\mbox{ implies that } K' \mbox{ is consistent }\}$.

\item $I_{hs}(K) = \min\{|H|\mid H\mbox{ is a hitting set for the interpretations of } K\}-1\mbox{, where }\min \varnothing =\infty$.

\item $I_C(K) = {\sf min} \{|{\sf Conflictbase}(i)| \mid i \in {\sf Models}(K)\}$
(where we are using 3VL).

\item $I_\eta(K)=1-\max\big\{\eta\in [0,1]\mid \{P(\phi)\geq \eta \mid \phi \in K\}\mbox{ is satisfiable}\big\}$.
\end{itemize}
\end{definition}

We explain the measures as follows.
$I_B$ is also called the drastic measure~\cite{HunterK08}: 
$0$ means consistent;
$1$ means inconsistent 
(it simply distinguishes between consistent and inconsistent KBs).
$I_M$ counts the number of minimal inconsistent subsets~\cite{HunterK08}.
$I_{\#}$ also counts the number of minimal inconsistent subsets, but
it gives larger sets a smaller weight
(the reason is that when a minimal inconsistent set contains more formulas than another minimal inconsistent set, the former is intuitively less inconsistent than the latter~\cite{HunterK08}).
$I_P$ counts the number of formulas that contribute essentially to one or more inconsistencies~\cite{GrantH11}.
$I_A$ uses maximal consistent subsets~\cite{GrantH11}.
Contradictory formulas are added as they do not
appear in any way in a maximal consistent set;
then $1$ must be subtracted to 
obtain $I_A(K) = 0$ for a consistent $K$ because every consistent knowledge base has a maximal consistent subset, namely $K$ itself.
$I_H$ counts the minimal number of formulas whose deletion makes the
set consistent~\cite{GrantH13}.
$I_{nc}$ uses the largest number such that all sets with that many
formulas are consistent~\cite{DoderRMO10}.
$I_{hs}$ uses the size of a minimal hitting set for the interpretations of the KB~\cite{Thimm16}.
$I_C$ counts the minimal number of atoms that must be assigned the
value B in a 3VL model~\cite{GrantH11}.
Finally, $I_\eta$ uses the PSAT concept~\cite{Knight02}: it finds the maximum probability lower bound $\eta$ that one can consistently assign to all formulas;
if $\eta$ is equal to $1$ then the KB is consistent.

In addition to devising many ways of measuring inconsistency,
researchers have also investigated properties that a good
inconsistency measure should possess.
These are called (rationality) postulates and we already gave
two of them: Consistency and Monotony, that all (absolute)
inconsistency measures should satisfy.
\cite{thimm18} lists 16 additional postulates but some of them are
oriented toward relative measures or deal with equivalent formulas
and so are not relevant for relational databases.
Thus, we will focus on the following postulates.

\begin{definition}[Postulates for Propositional Inconsistency Measures]
\label{def:postulates}
Let $K,K'$ be KBs, $\phi,\psi$ formulas, and $I$ and inconsistency measure. 
The postulates for inconsistency measures are as follows:
\begin{description}
\item[Free-Formula Independence]
If $\phi \in \Free(K)$, then $I(K) = I(K \setminus \{\phi\})$.

\item[Safe-Formula Independence]
If $\phi \in {\Safe}(K)$, then $I(K) = I(K \setminus \{\phi\})$.

\item[Penalty] 
If $\phi \in \Prob(K)$, then $I(K)>I(K\setminus\{\phi\})$.

\item[Dominance]
If $\phi$ is consistent and $\phi$ logically implies $\psi$, then
$I(K \cup \{\phi\}) \geq I(K \cup \{\psi\})$.

\item[Super-Additivity] 
If $ K\cap  K'=\varnothing$, then $I(K\cup  K') \geq I( K) + I( K')$.

\item[MI-Separability]  
If $\MI(K\cup K')=\MI(K)\cup \MI(K')$ and 
$\MI(K)\cap\MI(K')=\varnothing$, then $I(K\cup K')=I(K)+I(K')$.

\item[MI-Normalization]
If $M \in \MI(K)$, then $I(M) = 1$.

\item[Attenuation] 
If $M,M'\in \MI(K)$ 
and $|M|<|M'|$, then $I(M)>I(M')$.

\item[Equal Conflict] 
If $M,M'\in \MI(K)$
and $|M|=|M'|$, then $I(M)=I(M')$.

\item[Almost Consistency] 
If $M_1,M_2,\dots$ is a sequence of 
minimal inconsistent sets with $\lim\limits_{i\to \infty}|M_i|=\infty$, 
then $\lim\limits_{i\to \infty}I(M_i)=0$.

\end{description}
\end{definition}

The independence postulates mean that free (resp. safe) formulas
do not change the inconsistency measure. 
Penalty states that deleting a problematic formula decreases the measure.
Dominance deals with the case where a KB and two formulas $\phi$ and $\psi$
are given and $\phi$ is consistent and logically implies $\psi$.
Then the addition of $\psi$ to the KB cannot have a larger measure than the
addition of $\phi$.
Super-Additivity and MI-Separability give information about the
union of 2 KBs under certain conditions.
Super-Additivity deals with the case where the KBs are disjoint in which
case the measure of the union is at least as great as the sum of the measures
of the two KBs.
MI-Separability requires that the minimal inconsistent sets of the two
KBs partition the minimal inconsistent sets of the union in which case
the measure of the union is the sum of the measures of the two KBs.
MI-Normalization, Attenuation, Equal Conflict, and Almost Consistency
deal specifically with minimal inconsistent sets.
MI-Normalization requires every minimal inconsistent set to have measure $1$.
Attenuation requires larger size minimal inconsistent sets to have 
smaller measures; Equal Conflict requires minimal inconsistent set of
the same size to have the same measure.
Finally, Almost Consistency requires that as minimal inconsistent get larger the measures get closer and closer to $0$.

\subsection{Notation for Relational Databases}\label{sec:notationDBs}

We assume the existence of two finite (disjoint) sets: 
\relnames, the set of relation names, and \attnames, the set of attribute names.
We also assume a countably infinite database domain \dom, 
consisting of uninterpreted constants;
elements of the domain with different names are different elements.
Given a relation name $R\in  \relnames$, a \emph{relation scheme} for it 
is a sorted list $(A_1,\dots,A_n)$ of attributes $A_1,\dots,A_n \in \attnames$,
where $n$ is said to be the \textit{arity} of $R$ and
each attribute $A_i$ (with $i\in [1..n]$) has associated a domain $DOM(A_i)\subseteq \dom$. 
We use $R(A_1,\dots,A_n)$ to denote a relation scheme.
A \emph{database scheme} $\D$ is a nonempty finite set of relation schemes.
A \emph{tuple} over $R(A_1,\dots,A_n)$ is a mapping assigning to each attribute 
$A_i$ of $R$ a value $v_i\in DOM(A_i)$.
Given a tuple $\vec{t}=\langle v_1,\dots,v_n \rangle$, we use 
$\vec{t}[A_i]$ to denote the 
value $v_i$ of attribute $A_i$ of $\vec{t}$.
For an ordered list of attributes $\langle A_{i_1},\dots, A_{i_j} \rangle$, 
we use $\vec{t}[A_{i_1},\dots, A_{i_j}]$ to denote $\langle \vec{t}[A_{i_1}],\dots, \vec{t}[A_{i_j}]\rangle$.
A \emph{relation instance} (or simply relation) is a set of tuples over a given relation scheme,
and a \textit{database instance} (database) is a set of relations over a given database scheme.
A \emph{database instance} can be viewed as a finite Herbrand interpretation for
a (function-free) first-order language using constant symbols in \dom and 
predicate symbols in \relnames.
Hence, we write $R(v_1,\dots,v_n)$ or $R(\vec{t})$ for denoting the 
(ground) atom corresponding to the tuple $\vec{t}=\langle v_1,\dots, v_n \rangle$ over $R$.

Integrity constraints are first-order sentences expressing 
properties that are supposed to be satisfied by the database instance.
To define constraints, 
we extend the alphabet of the above-mentioned language to allow variables 
from a set \varnames of variables names (disjoint 
from \relnames\ and \attnames).
A \textit{term} is either a constant in \dom\ or a variable in \varnames.
An atom over a database scheme $\D$ is an expression of the form 
$R(\tau_1,\dots, \tau_n)$
where $R$ is a relation scheme in $\D$ having arity $n$ 
and $\tau_1,\dots, \tau_n$ are terms.

A \emph{denial constraint} over $\D$ is a first-order sentence of the form:\\
$
\forall\, \vec{x}_1,\dots,\vec{x}_k
\ [ \neg R_1(\vec{x}_1) \vee \cdots \vee \neg R_k(\vec{x}_k) \vee 
\varphi(\vec{x}_1,\dots,\vec{x}_k)]
$\\
where:
(\textit{i}) 
$\forall\, i\in [1..k]$, $\vec{x}_i$ are tuples of variables and
$R_i(\vec{x}_i)$ are atoms over $\D$; and
(\textit{ii})
$\varphi$ is a conjunction of built-in predicates of the form $\tau_i\circ \tau_j$
where $\tau_i$ and $\tau_j$ are 
variables in $\vec{x}_1,\dots, \vec{x}_k$ or constants, and $\circ\in\{=,\neq,>,<,\geq,\leq\}$.
In the following, 
we will omit the prefix of 
universal quantifiers and write  
$[ \neg R_1(\vec{x}_1) \vee \cdots \vee \neg R_k(\vec{x}_k) \vee 
\varphi(\vec{x}_1,\dots,\vec{x}_k)]$ for a denial constraint.
$k$ is said to be the \textit{arity} of the constraint.
Denial constraints of arity $2$ (resp. $3$) 
are called \textit{binary} (resp. \textit{ternary})
constraints.

A \textit{functional dependency} (FD) is a denial constraint
of the form: 
$
\neg [R(\vec{x},y,\vec{z}) \wedge R(\vec{x},u,\vec{w}) 
\wedge (y\neq u) ]
$
where $\vec{x},\vec{z},\vec{w}$ are tuples of variables.
It is usually written as
$R: X \rightarrow Y$ 
(or simply $X \rightarrow Y$ if the relation scheme is understood from the context),
where $X$ is the set of attributes of $R$ corresponding to $\vec{x}$ and
$Y$ is the attribute corresponding to $y$ (and $u$).

For a database scheme $\D$ and a set $\DC$ of integrity constraints over $\D$,
an instance $D$ of $\D$ is said to be \emph{consistent} w.r.t. $\DC$ iff 
$D\models \DC$ in the standard model-theoretic sense.

\section{Propositional Inconsistency Measures for Relational Databases}
\label{sec:extensions}

We first show how a relational database with constraints can be transformed
into a propositional KB where inconsistencies are mapped to.
This gives a systematic way to define the counterpart of existing propositional inconsistency measures in the context of relational databases:
applying measure $I$ to a database $D$
virtually means applying $I$ to the KB obtained by $D$ through the transformation.

The transformation process involves assigning a distinct propositional atom to each
tuple in the database and rewriting each denial constraint as a
propositional logic formula.

For the purpose of the transformation,
we write a database as a union of two distinct sets,
$DB = D \cup \DC$, where $D$ is the database instance and $\DC$ is the
set of constraints, denial constraints in our case.
So $|D|$ is the total number of tuples in all the relations of the database.
As usual, a set of formulas of $DB$ is a {\em minimal inconsistent subset}
if it is inconsistent and no proper subset is inconsistent.
A minimal inconsistent subset of $DB$ must contain an integrity constraint
and one or more tuples depending on the constraint.
We write $\MI(DB)$ for the set of minimal inconsistent subsets of $DB$.

Next we give the steps of the transformation.

\begin{definition}[Transformation]\label{def:trasformation}
The transformation from a relational database 
$DB=D \cup \DC$ to a propositional KB $K_{DB}$ is as follows.
\begin{itemize}
\item Let $A_D = \{a_1, \ldots, a_{|D|} \}$ be a set of $|D|$ 
propositional atoms.
\item Define a bijective function $f: D \rightarrow A_D$ 
that assigns a distinct propositional atom to each ground atom (i.e., tuple) in $D$.
\item Let ${\cal F}_D$ be the set of propositional formulas using $A_D$.
\item Define a partial function $g: \DC \rightarrow {\cal F}_D$ as follows:
For each constraint $c\in \IC$ of the form 
$c= 
[ \neg R_1(\vec{x}_1) \vee \cdots \vee \neg R_k(\vec{x}_k) \vee 
\varphi(\vec{x}_1,\dots,\vec{x}_k) ]
$,
if there is a sequence of tuples
$\langle \vec{t}_1, \ldots, \vec{t}_k \rangle$ whose substitution 
makes $c$ false, then 
$$
g(c) = \bigwedge_{\vec{t_i} \in T_{R_i}}
\ [\neg f(R_1(\vec{t}_1)) \vee \cdots \vee \neg f(R_k(\vec{t}_k)) 
\mid 
\varphi(\vec{t}_1,\dots,\vec{t}_k) \mbox{ is false for } 
1\leq i \leq k]
$$
where $T_{R_i}$ is the set of tuples in $D$ over $R_i$; otherwise
$g(c)$ is undefined.
\item 
We define 
$K_{DB}= \{ f(d) \mid d \in D\} \cup \{g(c) \mid c \in \DC \}$.
\end{itemize}
\end{definition}

The following example illustrates the definition.

\begin{example}
\label{ex:transform}
Consider the database scheme $\D_{ex}$ consisting of the 
relation scheme \textit{MealTicket}$($\textit{Number},\ \textit{Value},\ \textit{Holder},\ \textit{Date}$)$
whose instance contains the number, the value, the holder, 
and the issue date of meal tickets (one for each tuple) 
provided by a company to the employees.
The set $\DC_{ex}$ of integrity constraints consists of the following denial constraints:
\begin{itemize}
\item
$c_1= [ \neg$\textit{MealTicket}$(x_1,x_2,x_3,x_4) \vee x_2 > 0]$,
stating that the value (i.e., the amount of the ticket) 
of every tuple of \textit{MealTicket} must be a positive number.
\item
$c_2 = [ \neg$\textit{MealTicket}$(x_1,x_2,x_3,x_4) \vee \neg $\textit{MealTicket}$(x_1,x_5,x_6,x_7) \vee x_2 = x_5 ]$,
i.e., the FD 
\textit{Number}$\rightarrow $\textit{Value},
stating that there cannot be two distinct tickets 
with the same number and different values.
\item
$c_3 = [ \neg$\textit{MealTicket}$(x_1,x_2,x_3,x_4) \vee \neg $\textit{MealTicket}$(x_1,x_5,x_6,x_7) \vee x_3 = x_6 ]$,
i.e., 
\textit{Number}$\rightarrow $\textit{Holder}.
\item
$c_4 = [
\neg $\textit{MealTicket}$(x_1,x_2,x_3,x_4) \vee 
\neg $\textit{MealTicket}$(x_5,x_6,x_3,x_4) \vee
\neg $\textit{MealTicket}$(x_7,x_8,$ $x_3,x_4) \vee 
x_1 = x_5 \vee x_1 = x_7 \vee x_5 = x_7]$, 
stating the numerical dependency (see \cite{GM85TCS,GM85Iac})
\textit{Holder, Date} $\rightarrow^2 $\textit{Number} whose meaning 
is that for every holder and date there can be at most $2$ meal ticket numbers.
\end{itemize}
\begin{figure}[!t]
\begin{center}
\begin{tabular}{c||c|c|c|c||c}
  \hhline{~|t:====:t|~}
Atom & \raisebox{0cm}[3.2mm][1.1mm]{  
\textbf{\textit{Number}}}   & 
\textbf{\textit{Value}} & 
\textbf{\textit{Holder}} &
\textbf{\textit{Date}}& Tuple\\
  \hhline{~|:====:|}
$a_1$ & \raisebox{0cm}[3mm][1mm]{1001} &  15 & Matthew  & 2018-12-13 & $\vec t_1$\\
  \hhline{~||----||}
$a_2$ & \raisebox{0cm}[3mm][1mm]{1001}   & 15 & Matthew & 2018-12-18 & $\vec t_2$\\
  \hhline{~||----||}
$a_3$ & \raisebox{0cm}[3mm][1mm]{1001}  &  15 & Sophia  & 2018-12-17 & $\vec t_3$  \\
  \hhline{~||----||}
$a_4$ & \raisebox{0cm}[3mm][1mm]{1004}  &  20 & Sophia & 2018-12-17  & $\vec t_4$ \\
  \hhline{~||----||}
$a_5$ & \raisebox{0cm}[3mm][1mm]{1005}  &  0 & Alex & 2018-12-18  & $\vec t_5$ \\
  \hhline{~||----||}
$a_6$ & \raisebox{0cm}[3mm][1mm]{1006}  &  10 & Alex & 2018-12-18 & $\vec t_6$  \\
  \hhline{~||----||}
$a_7$ & \raisebox{0cm}[3mm][1mm]{1007}  &  20 & Alex & 2018-12-18  & $\vec t_7$ \\
\hhline{~|b:====:b|~}
\end{tabular}
\end{center}
\caption{Relation instance of \textit{MealTicket}.} 
\label{fig:mealTicketsInstance}
\end{figure}
Given the instance $D_{ex}$ of $\D_{ex}$ (see Figure~\ref{fig:mealTicketsInstance})
and $\DC_{ex}$, the transformation proceeds as follows:
\begin{enumerate}
\item 
We have $7$ propositional atoms in the set 
$A_{D_{ex}} = \{a_1, a_2, a_3, a_4, a_5, a_6, a_7 \}$.
\item
Each atom $a_i$ in $A_{D_{ex}}$ corresponds to a tuple $\vec t_i$ of \textit{MealTicket} 
as indicated in Figure~\ref{fig:mealTicketsInstance}.
This is defined by function $f$, for instance  
$f($\textit{MealTicket}$($1001, 15, Matthew, 2018-12-13 $)) = f(\vec t_1)= a_1$.
\item 
${\cal F}_{D_{ex}}$ is the set of propositional formulas using
the atoms in $A_{D_{ex}}$. 
\item
Every constraint in $\DC_{ex}$ is mapped by $g$ to a 
formula in ${\cal F}_{D_{ex}}$ as follows:
$g(c_1) = \neg a_5$,
$g(c_2)$ is undefined,
$g(c_3) = (\neg a_1 \vee \neg a_3) \wedge (\neg a_2 \vee \neg a_3)$,
$g(c_4) = \neg a_5 \vee \neg a_6 \vee \neg a_7$.
\item
Therefore, $DB_{ex}=D_{ex}\cup \DC_{ex}$ is transformed into the following  propositional KB:\\
$K_{DB_{ex}} 
=\{a_1, a_2, a_3, a_4, a_5, a_6, a_7, \neg a_5,
(\neg a_1 \vee \neg a_3) \wedge (\neg a_2 \vee \neg a_3),
\neg a_5 \vee \neg a_6 \vee \neg a_7\}$.
\end{enumerate}
\end{example}

\begin{proposition}
\label{prop:equivalent}
For every minimal inconsistent subset of $DB$ there is a unique
corresponding minimal inconsistent subset of $K_{DB}$.
\end{proposition}

The converse of the proposition does not hold because several
minimal inconsistent subsets of $DB$ may collapse to the same
inconsistent subset of $K_{DB}$ as we show in the following example.
\begin{example}
\label{ex:collapse}
Let $DB = \{R(0), \neg R(x) \vee x = 1, \neg R(x) \vee x = 2\}$.
Here $K_{DB} = \{a_1, \neg a_1\}$.
Then $\MI(DB) = 
\{ \{R(0), \neg R(x) \vee x = 1\}, \{ R(0), \neg R(x) \vee x = 2\}\}$,
while $\MI(K_{DB}) = \{\{a_1, \neg a_1\}\}$.
One tuple violates two integrity constraints; hence they are not
distinguished in $K_{DB}$.
\end{example}

We are now ready to formalize \textbf{propositional inconsistency measures 
for databases}.
For each inconsistency measure $I_x$ defined for propositional KBs we get 
an inconsistency measure for relational databases as $I_x(DB) = I_x(K_{DB})$.

\begin{example}
\label{ex:calculate}
Below are the results of calculating the inconsistency measures of the
relational database $DB_{ex}$ given in Example~\ref{ex:transform} by
calculating the inconsistency measures for $K_{DB_{ex}}$.

We use the fact that 
$
\MI(K_{DB_{ex}})=
\{
\{a_5,\neg a_5 \},
\{a_1, a_3, (\neg a_1 \vee \neg a_3) \wedge (\neg a_2 \vee \neg a_3)\},$ $
\{a_2, a_3, (\neg a_1 \vee \neg a_3) \wedge (\neg a_2 \vee \neg a_3)\},
\{a_5, a_6,a_7, \neg a_5 \vee \neg a_6 \vee \neg a_7 \}
\}
$.
\begin{itemize}
\item 
$I_B(DB_{ex}) = 1$ as $K_{DB_{ex}}$ is inconsistent.
\item 
$I_M(DB_{ex}) = 4$ as there are 4 minimal inconsistent subsets as given above 
(this is the total number of conjuncts in $\{g(c) \mid c \in \DC_{ex}\}$).
\item 
$I_{\#}(DB_{ex}) = \frac{1}{2} + \frac{1}{3} + \frac{1}{3} 
+ \frac{1}{4} = \frac{17}{12}$
as there are two minimal inconsistent subsets of size $3$ 
plus a minimal inconsistent subsets of size $2$ and one of size $4$.
\item 
$I_P(DB_{ex}) = 6 + 3 = 9$ as six atoms (i.e., tuples) plus three propositional formulas (i.e., constraints) are problematic, meaning that 
they are involved in an at least an inconsistency. 
\item 
$I_A(DB_{ex}) = 15 - 1 = 14$ as we next show by writing out all the
maximal consistent subsets (MCSs) according to the subset of the set of
the transformed integrity constraints that they contain.
In each case we write that subset first followed by the set of sets of
transformed tuples in the MCS.
\begin{itemize}
\item $\{ \neg a_5, (\neg a_1 \vee \neg a_3) \wedge (\neg a_2 \vee \neg a_3),
\neg a_5 \vee \neg a_6 \vee \neg a_7\}$.
There are $2$ MCSs that contain all $3$ transformed
constraints: one also has
$\{a_1, a_2, a_4, a_6, a_7\}$ while the other one has 
$\{a_3, a_4, a_6, a_7\}$.
\item $\{ \neg a_5, \neg a_5 \vee \neg a_6 \vee \neg a_7\}$.
There is $1$ MCS that contains exactly those integrity
constraints: it has 
$\{a_1, a_2, a_3, a_4, a_6, a_7\}$.
\item $\{ (\neg a_1 \vee \neg a_3) \wedge (\neg a_2 \vee \neg a_3),
\neg a_5 \vee \neg a_6 \vee \neg a_7\}$.
There are $6$ MCSs that contain exactly those integrity
constraints: namely those that also contain 
$\{a_1, a_2, a_4, $ $a_5, a_6\}$,
$\{a_1, a_2, a_4, a_5, a_7\}$,
$\{a_1, a_2, a_4,$ $a_6, a_7\}$,
$\{a_3, a_4, a_5, a_6\}$,
$\{a_3, a_4, a_5, a_7\}$,
$\{a_3, a_4, a_6, a_7\}$,
\item $\{ (\neg a_1 \vee \neg a_3) \wedge (\neg a_2 \vee \neg a_3)\}$.
There are $2$ MCSs that contain exactly this integrity
constraints: namely those that also contain 
$\{a_1, a_2, a_4, a_5, a_6, a_7\}$ and $\{a_3, a_4, a_5, a_6, a_7\}$.
\item $\{\neg a_5 \vee \neg a_6 \vee \neg a_7\}$.
There are $3$ MCSs that contain exactly this integrity
constraints: namely those that also contain 
$\{a_1, a_2, a_3,a_4, a_5, a_6\}$, 
$\{a_1, a_2, a_3, a_4, a_5, a_7\}$,
and $\{a_1, a_2, a_3, a_4, a_6, a_7\}$,
\item $\emptyset$. There is $1$ MCS with no integrity constraints:
$\{a_1, a_2, a_3,a_4, a_5, a_6, a_7\}$.
\item
Finally, there are no MCSs containing either 
$\{ \neg a_5, (\neg a_1 \vee \neg a_3) \wedge (\neg a_2 \vee \neg a_3)\}$
or $\{ \neg a_5\}$ as the set of transformed constraints 
because, for any consistent
subset containing one of these sets of constraints, the addition of 
$\neg a_5 \vee \neg a_6 \vee \neg a_7$ does not violate consistency.
\end{itemize}
\item 
$I_H(DB_{ex}) = 2$
as the set $\{a_3,a_5\}$ having cardinality $2$ 
intersects with any minimal inconsistent subset 
(notice that also  
$\{a_5, (\neg a_1 \vee \neg a_3) \wedge (\neg a_2 \vee \neg a_3)\}$
could be used to get the same value).
\item  
$I_{nc}(DB_{ex}) = 10 - 1 = 9$ as the set $\{a_5, \neg a_5\}$ has size
$2$ and is inconsistent.
\item 
$I_{hs}(DB_{ex}) = 1$ as in our setting it gets the same value of $I_B$---see Proposition~\ref{prop:hs=B} below.
\item 
$I_C(DB_{ex}) = 2$ as there is a 3VL-model assigning
B to $a_3$ and $a_5$ (and T to the other atoms).
\item 
$I_\eta(DB_{ex}) = 0.5$ because of the presence of both $a_5$ and
$\neg a_5$. 
Let $Int(a_5)$ be the set of interpretations for which $a_5$ is true,
and $Int(\neg a_5)$ the set of interpretations for which $a_5$ is false.
A probability function $\pi$ such that 
$\sum_{i \in Int(a_5)} \pi(i) =0.5$ and $\sum_{i \in Int(\neg a_5)} \pi(i) = 0.5$
gives the highest probability, $0.5$, for both formulas.
\end{itemize}
\end{example}

In the propositional case, of the 10 measures we presented no 2
measures give the same result for all KBs.
But because of the special structure of the $K_{DB}$s, the
hitting set measure and the drastic measure give identical results.

\begin{proposition}\label{prop:hs=B}
For all $DB$s, $I_{hs}(DB) = I_B(DB)$.
\end{proposition}

\subsection{Rationality Postulates Satisfaction for Relational Databases}

\cite{thimm18} includes a list that for each inconsistency measure
shows for each postulate whether or not it is satisfied.
The results for the satisfaction of postulates carries over to
relational databases by the transformation we presented.
However, as the transformed relational database contains only a
restricted set of formulas, for some of the measures additional
postulates are also satisfied.
Before we can get to this we need to fix the meaning of 
some terminology in our context.
We say that a formula $R(\vec{t})$ in $DB$ is {\em free} (resp. {\em safe})
if $f(R(\vec{t}))$ is free (resp. safe) in $K_{DB}$.
Also, a constraint $c$ in $DB$ is {\em free} (resp. {\em safe}) in $DB$
if $g(c)$ is undefined.
A formula that is not free is {\em problematic}.

We start by proving a simple result about free and safe formulas.

\begin{proposition}
\label{prop:freesafe}
A formula in $DB$ is free iff it is a tuple that is not part of any
inconsistency or an integrity constraint that is not violated. 
Furthermore, every free formula is safe.
\end{proposition}

Therefore, free and safe formulas are identical,
from which the following corollary follows.

\begin{corollary}
\label{cor:freesafe}
An inconsistency measure for $DB$ satisfies Free-Formula Independence iff it
satisfies Safe-Formula Independence.
\end{corollary}

Several postulates (e.g., Penalty) deal with what happens to the inconsistency
measure when a formula is deleted.
The point is that the formula is deleted from $DB$, not $K_{DB}$.
Hence we must determine how such a deletion affects $K_{DB}$.
$DB$ contains ground atoms representing tuples in relations and
constraints.
When a ground atom $R(\vec{t})$ is deleted from $DB$, its transformation,
the propositional atom $f(R(\vec{t}))$ must be deleted from $K_{DB}$.
But that is not all. The propositional atom may also appear in some
conjuncts of formulas transformed by $g$ from constraints.
All those conjucnts must be deleted as well. 
For example, if 
$K_{DB} = \{a_1, a_2, a_3, (\neg a_1 \vee \neg a_2) \wedge (\neg a_1 \vee
\neg a_3) \}$
then the deletion of $R(\vec{t})$, where $f(\vec{t}) = a_2$ 
requires the deletion of both $a_2$ and $\neg a_1 \vee \neg a_2$.
So $K_{DB \setminus \{R(\vec{t})\}} = \{a_1, a_3, \neg a_1 \vee \neg a_3\}$.
The deletion of an integrity constraint $c$ is done as follows.
If $g(c)$ is undefined then $K_{DB \setminus \{c\}} = K_{DB}$.
If there is an integrity constraint $c' \neq c$ such that
$g(c) = g(c')$ then $K_{DB \setminus \{c\}} = K_{DB}$;
otherwise $K_{DB \setminus \{c\}} = K_{DB} \setminus \{g(c)\}$.

Some postulates (e.g. Super-Additivity) refer to the union of two $DB$s.
We assume that they have the same schema.
Let $DB_1 = D_1 \cup \DC_1$ and
$DB_2 = D_2 \cup \DC_2$.
In general, it is not the case that 
$K_{DB_1 \cup DB_2} = K_{DB_1} \cup K_{DB_2}$.
For example, a functional dependency in $\DC_1$ may apply to tuples in 
$D_2$ or a mixture of tuples from $D_1$ and $D_2$.
Also, in the transformations we start with the atoms $a_1, a_2, \ldots$
for both databases but the $a_1$ obtained from $D_1$ need not be
the same as the $a_1$ obtained from $D_2$.
So we must take the union of the databases first and then do the transformation.
For example, let $R(B,E,H)$ be a relation scheme, if 
$D_1 = \{R(b_1,e_1,h_1), R(b_1,e_2, h_2)\}$, 
$\DC_1 = \{B \rightarrow E\}$, and 
$D_2 = \{R(b_1,e_3,h_2)\}$, 
$\DC_2 = \{B \rightarrow E, H \rightarrow E\}$, then 
$K_{DB_1} = \{a_1, a_2, \neg a_1 \vee \neg a_2\}$ and
$K_{DB_2} = \{a_1\}$. 
Taking the union of the databases we get 
$D_1 \cup D_2 = \{R(b_1,e_1,h_1), R(b_1,e_2, h_2), R(b_1,e_3,h_2)\}$, 
$\DC_1 \cup \DC_2 = \{ B \rightarrow E,  H \rightarrow E\}$.
Hence, 
$K_{DB_1 \cup DB_2} = 
\{a_1, a_2, a_3, (\neg a_1 \vee \neg a_2) \wedge 
(\neg a_1 \vee \neg a_3) \wedge (\neg a_2 \vee \neg a_3),
\neg a_2 \vee \neg a_3\}$
that cannot be obtained strictly from $K_{DB_1}$ and $K_{DB_2}$.

A similar situation occurs for intersection, which used for instance in 
Super-Additivity.
That is, for 
$DB_1 = D_1 \cup \DC_1$ and $DB_2 = D_2 \cup \DC_2$,
in order to compute $K_{DB_1 \cap DB_2}$ we must first obtain
$D_1 \cap D_2$ and $\DC_1 \cap \DC_2$ and then proceed with the
transformation. 

Finally, consider how a formula in $DB$ may logically imply 
another formula (this is considered by Dominance).
Clearly, both must be integrity constraints, say $c \rightarrow c'$
where
$c = [ \neg R_1(\vec{x}_1) \vee \cdots \vee \neg R_k(\vec{x}_k) \vee 
\varphi(\vec{x}_1,\dots,\vec{x}_k)]$.
There are two cases.
One case is where $c'$ contains one or more additional disjuncts, say
$c' = [ \neg R_1(\vec{x}_1) \vee \cdots \vee \neg R_k(\vec{x}_k) \vee 
\neg R_{k+1}(\vec{x}_{k+1}) \cdots \neg R_m(\vec{x}_m) \vee
\varphi(\vec{x}_1,\dots,\vec{x}_k)]$.
Another case is $c' = [ \neg R_1(\vec{x}_1) \vee \cdots \vee \neg R_k(\vec{x}_k) \vee 
\varphi'(\vec{x}_1,\dots,\vec{x}_k)]$ 
where $\varphi(\vec{x}_1, \ldots, \vec{x}_k) \rightarrow
\varphi'(\vec{x}_1, \ldots, \vec{x}_k)$.
Clearly, it is also possible to have a combination of these cases.

Next we present our main result concerning the postulates
that inconsistency measures satisfy when restricted to relational
databases.

\begin{theorem}
\label{th:satisfy}
The satisfaction of postulates for propositional inconsistency measures for 
databases is as given in Table~\ref{tab:postulates1}.
\end{theorem}

\begin{table}[h!]
{
\centering
\begin{tabular}{||c||c|c|c|c|c|c|c|c|c|c||} 
\hhline{~|t:==========:t|}
\multicolumn{1}{c||}{} & \multicolumn{10}{c||}{\bf Propositional Inconsistency Measures for Databases} \\
\hhline{~|:==========:|}
\multicolumn{1}{c||}{} & \hspace*{2mm}$I_B$\hspace*{2mm} &\hspace*{2mm} $I_M$\hspace*{2mm} &\hspace*{2mm} $I_{\#}$ \hspace*{2mm}&\hspace*{2mm} $I_P$\hspace*{2mm} & \hspace*{2mm}$I_A$ \hspace*{2mm}& \hspace*{2mm}$I_{H}$\hspace*{2mm} &\hspace*{2mm} $I_{nc}$\hspace*{2mm} & $I_{hs}$\hspace*{2mm} & \hspace*{2mm}$I_{C}$\hspace*{2mm} & \hspace*{2mm}$I_{\eta}$  \\ 
\hhline{|t:===========||}
Free-Formula Independence &\cmark   &\cmark    &\cmark    & \cmark    &\cmark     & \cmark   &\xmark    &\cmark     &\cmark        & \cmark   \\
\hhline{||-||----------||}
Safe-Formula Independence &\cmark    &\cmark    &\cmark    &\cmark     &\cmark     &\cmark   &\xmark    & \cmark    &\cmark        & \cmark   \\
\hhline{||-||----------||}
Penalty &\xmark    &\cmark    &\cmark    &\cmark     &{\color{blue}\cmark$^*$}    &\xmark    & \cmark    & \xmark   & \xmark   &\xmark    \\
\hhline{||-||----------||}
Dominance &\cmark    &\xmark    &\xmark    & \xmark    &\xmark    &\xmark     &\xmark    & \cmark     & \cmark       &\cmark    \\ 
\hhline{||-||----------||}
Super-Additivity &\xmark    &\cmark    &\cmark    &\cmark     &{\color{blue}\cmark$^*$}    &\cmark    &\cmark    & \xmark    &{\color{blue}\cmark$^*$}    &\xmark    \\
\hhline{||-||----------||}
MI-Separability &\xmark   &\cmark    &\cmark    &\xmark     &\xmark    &\xmark     & \xmark   & \xmark   & \xmark   &  \xmark  \\
\hhline{||-||----------||}
MI-Normalization & \cmark   &\cmark    &\xmark    &\xmark     &\xmark    &\cmark    &\cmark     & {\color{blue}\cmark$^*$}    &{\color{blue}\cmark$^*$}    & \xmark   \\
\hhline{||-||----------||}
Attenuation &\xmark  &\xmark    &\cmark    &\xmark     &\xmark    &\xmark     &\xmark    &\xmark     & \xmark   & \cmark   \\
\hhline{||-||----------||}
Equal Conflict &\cmark    &\cmark    &\cmark    &\cmark     &\cmark     &\cmark    &\cmark     &\cmark     &{\color{blue}\cmark$^*$}   & \cmark   \\
\hhline{||-||----------||}
Almost Consistency&\xmark   &\xmark    &\cmark    &\xmark   &\xmark    & \xmark    &\xmark    &\xmark    &\xmark    & \cmark   \\
\hhline{|b:===========:b|}
\end{tabular}
\vspace*{2mm}
\caption{Postulates satisfaction for propositional inconsistency measures 
applied to databases via the transformation.
\cmark: satisfied for databases because satisfied for KBs.
{\color{blue}\cmark$^*$}: satisfied for databases but not for KBs.
\xmark: not satisfied for databases (hence not satisfied for KBs).}
\label{tab:postulates1}
}
\end{table}

\section{Measuring Inconsistency by Blaming Database Tuples Only}\label{sec:second-method}

In the previous section, we transformed a relational database to a
propositional knowledge base and used previous studies of inconsistency
measures for propositional knowledge bases to obtain the corresponding relational 
database inconsistency measures. 
In this section, we propose a different method:
we develop inconsistency measures directly for relational databases
in analogy with the propositional case but without doing a transformation.
Furthermore, we assume that a fixed database schema $\D$ and set $\DC$ of integrity constraints are given that are used for all the databases.
Hence our measures will be calculated using only database tuples (with the
integrity constraints in the background).

We start with the basic definitions needed to define inconsistency
measures and their properties in this context.
We will omit the (fixed set of) integrity constraints in the terminology.
In order to distinguish from inconsistency measures 
obtained from the transformation, where $I$ is used, here we use $\I$,
again with subscripts.
This means that a database $D$ now is simply a set of relational tuples, and 
$\IC$ is used only for determining the consistent and inconsistent subsets of $D$.
Thus, in contrast with the case of propositional knowledge bases 
where all formulas have equal status, 
here formulas representing integrity constraints are regarded as not faulty: 
they do not belong to minimal inconsistent subsets and thus cannot be problematic.

A {\bf minimal inconsistent subset} of $D$ is a set of tuples $X$ such
that $X$ is inconsistent (with respect to $\IC$) and no proper subset
of $X$ is inconsistent.
As before we denote by $\MI(D)$ the set of minimal inconsistent subsets.
Similarly, a {\bf maximal consistent subset} is a set of tuples $Y$ that
is consistent and no proper superset of $Y$ is consistent.
We write $\MC(D)$ for the set of maximal consistent subsets (of $D$).
Any tuple that occurs in a minimal inconsistent subset is {\bf
problematic}; otherwise it is {\bf free}.
We use $\Prob(D)$ and $\Free(D)$ to denote the sets of problematic and free tuples of $D$.
Although no relational tuple by itself is inconsistent, it is 
possible to have a minimal inconsistent subset with a single element,
because of the way inconsistency is defined with respect to $\IC$.
We call such a tuple a {\bf contradictory} tuple and write
$\Contradictory(D)$ for the set of contradictory tuples.

If we deal only with the database tuples, 
and use integrity constraints only for check consistency, 
there is no counterpart to the concept of an interpretation 
that assigns a truth value to each atom (i.e., tuple).
So we start by providing inconsistency measures that do not rely 
on the concept of interpretation in Definition~\ref{def:seven}, and 
later separately define the counterparts for $I_{hs}$, $I_C$, and $I_\eta$.
The concept of safe formula also has no counterpart, so we will not
deal with Safe-Formula Independence separately.
Additionally, it is not possible for one relational tuple to logically imply
another so Dominance applies only to the case where the two formulas
added are identical.
Consequently, Dominance is always satisfied (trivially) and we will
omit it from consideration.
Furthermore, in considering several postulates, we do not have to do
extra work to define union and intersection as was needed for the
translated version.

We now write the definitions of the inconsistency
measures and postulates in this new framework.
We write $D$ for an arbitrary relational database (instance) and
${\cal D}$ for the set of all databases (for a predefined schema,
domains, and integrity constraints).

\begin{definition}
\label{def:incymeas2}
A function $\I: {\cal D} \rightarrow \mathbb{R}^{\geq 0}_\infty$ is an 
{\bf inconsistency measure} 
if the following two conditions hold for all $D,D'\in\cal{D}$:
\begin{description}

\item[Consistency] $\I(D) = 0$ iff $D$ is consistent.

\item[Monotony] If $D\subseteq D'$, then $\I(D)\leq \I(D')$.
\end{description}
\end{definition}

\begin{definition}[Database Inconsistency Measures]
\label{def:seven}
For a database $D$, the inconsistency measures 
$\I_B$, $\I_M$, $\I_{\#}$, $\I_P$, $\I_A$, $\I_H$, and $\I_{nc}$
are such that
\begin{itemize}
\item 
$\I_B(D) = 1$ if $D$ is inconsistent and $\I_B(D) = 0$ if $D$ is consistent.
\item 
$\I_M(D) = |\MI(D)|$.
\item
$\I_{\#}(D) = 
\left\{
\begin{array}{ll} 
0 &\hspace*{2mm} \mbox{if } D \mbox{ is consistent},\\
\sum\limits_{X \in {\MI}(D)} \frac{ 1 }{ | X | } &\hspace*{2mm} \mbox{otherwise} .
\end{array} 
\right. $
\item 
$\I_P(D) = | \Prob(D)  |$.
\item 
$\I_A(D) = (| \MC(D) | + | \Contradictory(D) |) -1$.
\item 
$\I_H(D) = \min\{ |X|\ s.t.\ X \subseteq D \mbox{ and } \forall M \in \MI(D),\
X \cap M \neq \emptyset \}$.
\item 
$\I_{nc}(D)= |D|-\max\{\, n\mid \forall D'\subseteq D: |D'|=n\mbox{ implies } D' \mbox{ is consistent}\}$. 
\end{itemize}
\end{definition}

To define the counterparts of measures $I_{hs}$ and $I_C$ that strictly rely on the concept of interpretation of the underlying knowledge base, 
we leverage on $K_{DB}$, with $DB=D \cup \DC$, of Definition~\ref{def:trasformation}.
However, in this case, we will formally require that the (formulas encoding the) 
integrity constraints must hold in every interpretation of $K_{DB}$, 
meaning that some (formulas encoding) tuples may be assigned $F$ for a classical 
two-valued interpretation or $B$ for a 3VL interpretation of $K_{DB}$.

We use $\mathbb{I}_C$ to denote the set of such two-valued 
interpretations for $K_{DB}$,
that is, $\mathbb{I}_C=\{ i \mid i$ is a classical interpretation for $K_{DB}$
and $\forall c\in \IC,\ i(g(c))=T\}$.

Moreover, we use $\mathbb{M}_{3VL}$ to denote the set of 
3VL models of $K_{DB}$ such that no atom (i.e., tuple) 
is assigned \textit{false} 
and every other formula  (i.e., integrity constraint) is assigned \textit{true}:\\ $\mathbb{M}_{3VL}=
\{ i \mid i$ is a 3VL model for $K_{DB}$ s.t. $\forall t\in D$, $i(f(t))$ is either $T$ or $B$ and $\forall c \in \IC,\ i(g(c)) = T \}$.

Using $\mathbb{I}_C$ and $\mathbb{M}_{3VL}$ we can formally define measures $\I_{hs}$ and $\I_C$ that, likewise the measures of Definition~\ref{def:seven}, measure 
inconsistency in terms of database tuples only.

\begin{definition}[Measures $\I_{hs}$, $\I_{C}$]
\label{def:incmeas-two}
For a database $D$, the inconsistency measures 
$\I_{hs}$ and $\I_C$ are such that
\begin{itemize}
\item $\I_{hs}(D) = \min\{\, |H|\ s.t.\ H$ is a hitting set for $\mathbb{I}_C\} - 1$, 
where $\min \emptyset =\infty$.

\item $\I_C(D) = {\sf min} \{|{\sf Conflictbase}(i)| \mid i \in \mathbb{M}_{3VL}\}$.
\end{itemize}
\end{definition}

Finally, we define the counterpart of the probabilistic measure $I_{\eta}$
by forcing the formulas of $K_{DB}$ representing integrity constraints to
be assigned a probability equal to 1, i.e., they are not relaxed
as done for probabilistic databases with integrity constraints~\cite{FlescaFP14}. 
Formally, given a database $D$ and a set of integrity constraints $\IC$,
we define the PSAT instance 
$\Gamma_{\IC}(D)=\{P(f(t))\geq \eta \mid t\in D\}\cup 
\{P(g(c))= 1 \mid c\in \IC\}$,
which enables the following definition of inconsistency measure.

\begin{definition}[Measure $\I_{\eta}$]
\label{def:psat-based-measure}
Given a database $D$ and a set of integrity constraints $\IC$, 
the inconsistency measure $\I_\eta$ is such that
$\I_\eta(D)=1-\max\big\{\eta\in [0,1]\, \mid\, \Gamma_{\IC}(D)\mbox{ is satisfiable}\big\}$.
\end{definition}

Thus, $\I_\eta(D)$ is one minus the maximum probability lower bound one 
can consistently assign to all tuples in $D$.

Next we give the definitions for the postulates.

\begin{definition}[Postulates for Database Inconsistency Measures]
\label{def:dbpostulates}
Let $D,D'$ be databases, $R(\vec{t})$ a tuple of $D$,
and $\I$ an inconsistency measure. 
The postulates for database inconsistency measures are as follows:
\begin{description}
\item[Free-Formula Independence]
If $R(\vec{t}) \in \Free(D)$, then $\I(D) = \I(D \setminus \{R(\vec{t})\})$.

\item[Penalty] 
If $R(\vec{t}) \in \Prob(D)$, then $\I(D)>\I(D\setminus\{R(\vec{t})\})$.

\item[Super-Additivity] 
If $ D\cap  D'=\varnothing$, then $\I(D\cup  D') \geq \I( D) + \I( D')$.

\item[MI-Separability]  
If $\MI(D\cup D')=\MI(D)\cup \MI(D')$ and 
$\MI(D)\cap\MI(D')=\varnothing$, then $\I(D\cup D')=\I(D)+\I(D')$.

\item[MI-Normalization]
If $M \in \MI(D)$, then $\I(M) = 1$.

\item[Attenuation] 
If $M,M'\in \MI(D)$ 
and $|M|<|M'|$, then $\I(M)>\I(M')$.

\item[Equal Conflict] 
If $M,M'\in \MI(D)$
and $|M|=|M'|$, then $\I(M)=\I(M')$.

\item[Almost Consistency] 
If $M_1,M_2,\dots$ is a sequence of 
minimal inconsistent sets with $\lim\limits_{i\to \infty}|M_i|=\infty$, 
then $\lim\limits_{i\to \infty}\I(M_i)=0$.

\end{description}
\end{definition}

Now we go back to Example~\ref{ex:transform} and calculate the
values for the database inconsistency measures.

\begin{example} 
\label{ex:dbcalculate}
In accordance with our notation, we write $D_{ex}$ for the $7$
relational tuples $\vec t_1,\dots,\vec t_7$ of Figure~\ref{fig:mealTicketsInstance}. 
We use the fact that 
$\MI(D_{ex})=
\{
\{\vec t_5 \},
\{\vec t_1, \vec t_3\},
\{\vec t_2, \vec t_3\}
\}
$.
\begin{itemize}
\item 
$\I_B(D_{ex}) = 1$ as the database is inconsistent.
\item 
$\I_M(D_{ex}) = 3$ as there are 3 minimal inconsistent subsets as given above. 
\item 
$\I_{\#}(D_{ex}) = 1 + \frac{1}{2} + \frac{1}{2} = 2$
as there is one minimal inconsistent subsets of size 
$1$ and two minimal inconsistent subsets of size $2$.
\item 
$\I_P(D_{ex}) = 4$ as there are $4$ distinct tuples in $\MI(D_{ex})$.
\item 
$\I_A(D_{ex}) = 2$ because there are $2$ maximal consistent subsets:
$\{\vec t_1,\vec t_2,\vec t_4, \vec t_6, \vec t_7 \}$ and $\{\vec t_3,\vec t_4, \vec t_6, \vec t_7\}$, and one contradictory tuple: $\vec  t_5$.
\item 
$\I_H(D_{ex}) = 2$
as the set $\{\vec  t_3, \vec t_5 \}$
intersects with each minimal inconsistent subset. 
\item  
$\I_{nc}(D_{ex}) = 7 - 0 = 7$ as the set $\{\vec t_5\}$
has size $1$ and is inconsistent.
\item 
$\I_{hs}(D_{ex}) = \infty$ because of the contradictory tuple $\vec{t}_5$.
\item 
$\I_C(D_{ex}) = 2$ as in the database setting it gives the same value of $\I_H$---see Proposition~\ref{prop:C=HforDatabases}.
\item 
$\I_\eta(D_{ex}) = 1$ because the maximum probability that can be assigned to 
tuple $\vec t_5$ is zero.
\end{itemize}
\end{example}

It turns out that unlike for the case of propositional inconsistency measures, 
$\I_{hs}(D) = \I_B(D)$ does not hold.
However, for database inconsistency measures there is an equality that does hold: 
$\I_{C}(D) = \I_H(D)$.

\begin{proposition}\label{prop:C=HforDatabases}
For any database $D$, 
$\I_{C}(D) = \I_H(D)$.
\end{proposition}

We now state the result of postulates satisfaction for database inconsistency measures.
It turns out that the satisfaction of the postulates for database
inconsistency measures is very similar to but not identical to
the satisfaction of the corresponding postulates for the propositional
inconsistency measures.
However, in this case the satisfaction results for propositional KBs
cannot be used.

\begin{theorem}
The satisfaction of postulates for database inconsistency measures
is as given in Table~\ref{tab:postulates2}.
\end{theorem}
\begin{table}[h!]
{
\centering
\begin{tabular}{||c||c|c|c|c|c|c|c|c|c|c||} 
\hhline{~|t:==========:t|}
\multicolumn{1}{c||}{} & \multicolumn{10}{c||}{\bf Database Inconsistency Measures} \\
\hhline{~|:==========:|}
\multicolumn{1}{c||}{} & \hspace*{1mm}$\I_B$ \hspace*{1mm}& \hspace*{1mm}$\I_M$\hspace*{1mm} &\hspace*{1mm} $\I_{\#}$\hspace*{1mm} & $\I_P$\hspace*{1mm} & \hspace*{1mm}$\I_A$\hspace*{1mm} & \hspace*{1mm}$\I_{H}$\hspace*{1mm} & \hspace*{1mm}$\I_{nc}$\hspace*{1mm} & $\I_{hs}$\hspace*{1mm} &\hspace*{1mm} $\I_{C}$\hspace*{1mm} &\hspace*{1mm} $\I_{\eta}$  \\ 
\hhline{|t:===========||}
Free-Formula Independence &\cmark &\cmark &\cmark &\cmark &\cmark &\cmark &\xmark &\cmark &\cmark &\cmark    \\
\hhline{||-||----------||}
Safe-Formula Independence &\cmark  &\cmark &\cmark &\cmark &\cmark &\cmark &\xmark &\cmark &\cmark & \cmark   \\
\hhline{||-||----------||}
Penalty &\xmark &\cmark &\cmark &\cmark &\textcircled{\xmark} &\xmark &\cmark &\xmark &\xmark & \xmark   \\
\hhline{||-||----------||}
Dominance &\cmark &\textcircled{\cmark} &\textcircled{\cmark} &\textcircled{\cmark} &\textcircled{\cmark} &\textcircled{\cmark} &\textcircled{\cmark} &\cmark &\cmark &\cmark    \\ 
\hhline{||-||----------||}
Super-Additivity &\xmark &\cmark &\cmark &\cmark &\cmark &\cmark &\cmark &\xmark &\cmark &\xmark    \\
\hhline{||-||----------||}
MI-Separability &\xmark &\cmark &\cmark &\xmark &\xmark &\xmark &\xmark &\xmark &\xmark &\xmark    \\
\hhline{||-||----------||}
MI-Normalization & \cmark &\cmark &\xmark &\xmark &\xmark &\cmark &\cmark &\textcircled{\xmark} &\cmark &\xmark    \\
\hhline{||-||----------||}
Attenuation &\xmark &\xmark &\cmark &\xmark &\xmark &\xmark &\xmark &\xmark &\xmark &\xmark    \\
\hhline{||-||----------||}
Equal Conflict &\cmark &\cmark &\cmark &\cmark &\cmark &\cmark &\cmark &\cmark &\cmark &\cmark    \\
\hhline{||-||----------||}
Almost Consistency&\xmark &\xmark &\cmark &\xmark &\xmark &\xmark &\xmark &\xmark &\xmark & \cmark   \\
\hhline{|b:===========:b|}
\end{tabular}
\vspace*{2mm}
\caption{Postulates satisfaction of database inconsistency measures.
Results different from those for the corresponding measure in Table~\ref{tab:postulates1} are highlighted by circles.
}
\label{tab:postulates2}
}
\end{table}

Before starting our complexity analysis of inconsistency measures,
we note that measure $\I_{nc}$ becomes not useful in the database setting as it 
will always gives $|D|-1$ if there is any functional dependency violation.
So we will no longer consider it in what follows.

\subsection{Complexity of Database Inconsistency Measures}\label{sec:complexity}

We investigate the data-complexity~\cite{ChandraH80,Vardi82} of the following three decision problems,
which intuitively ask if a given constant value $v$  is, respectively, 
lower than, greater than, or equal to the value returned by a given inconsistency measure when applied to a given database.

\begin{definition}[Lower Value (\LV), Upper Value (\UV), and Exact Value (\EV) problems]\label{def:decision-problems}
Let $\I$ be an inconsistency measure.
Given a database $D$
over a fixed database scheme with a fixed set of constraints, and 
a positive value $v\in \R^{>0}$,
\LV$_{\I}(D,v)$ is the problem of deciding whether $\I(D)\geq v$.\\
Given $D$ and a non-negative value $v'\in \R^{\geq 0}$,
\UV$_{\I}(D,v)$ is the problem of deciding whether $\I(D)\leq v'$, and
\EV$_{\I}(D,v)$ is the problem of deciding whether $\I(D)= v'$.
\end{definition}

\begin{table}[t!]
\begin{center}
\begin{tabular}{||l||c|c|c||c||}
\hhline{|t:=====:t|}
Inconsistency Measure(s) & \LV$_{\I}(D,v)$ & \LV$_{\I}(D,v)$ & \EV$_{\I}(D,v)$& \IM$_{\I}(D)$\\
\hhline{t|:=====||}
$\I_B$,$\I_M$,$\I_\#$,$\I_P$ & $P$ & $P$ & $P$ & $FP$\\
\hhline{||-----||}
$\I_A$ & $CP$ &  $CP$ &  $CP$ & $\#P$-complete$^*$\\
\hhline{||-----||}
$\I_H$, $\I_C$ & $coNP$-complete  & $NP$-complete & $D^p$-complete & $FP^{NP[log\ n]}$-complete\\
\hhline{||-----||}
$\I_\eta$ & $coNP$-complete  & $NP$-complete  & $D^p$
&$FP^{NP}$\\
\hhline{|b:=====:b|}
\end{tabular}
\end{center}
\caption{Complexity of Lower Value (\LV), Upper Value (\UV), 
Exact Value (\EV), and Inconsistency Measurement (\IM) problems.
($^*$: $\#P$-hardness follows from a result in~\cite{LivshitsK17},
which also implies that \IM, as well as \LV, \UV, and \EV, 
are polynomial for \textit{chain} FD schemas).}
\label{tab:summaryRepair}
\end{table}

We also consider the function problem of determining the value of an inconsistency measure.

\begin{definition}[Inconsistency Measurement (\IM) problem]\label{def:function-problem}
Let $\I$ be an inconsistency measure.
Given a database $D$
over a fixed database scheme with a fixed set of constraints, 
\IM$_{\I}(D)$ is the problem of computing the value of $\I(D)$.
\end{definition}

The following theorem characterizes the complexity of the database inconsistency measures.
We leave the investigation of the complexity of $\I_{hc}$ to future work.

\begin{theorem}
The complexity of the database inconsistency measures is as given in Table~\ref{tab:summaryRepair}.
\end{theorem}

\section{Conclusions and Future Work}\label{sec:conclusionAndFW}

Inconsistency in databases is not the exception, it is quite common.
Quantifying and monitoring the amount of inconsistency in a database 
helps to get information on the health status of data,
whose quality is more and more important nowadays---the global market of 
data quality tools is 
expected to grow from USD 610.2 Million in 2017 to USD 1,376.7 Million by 2022~\cite{marketsAndmarkets}.

In this paper, we have taken the first steps towards a formal framework for 
measuring inconsistency in databases.
We believe that the definition and investigation of inconsistency measures for databases 
benefit from our systematic approach to the problem, 
which stems from what has been done in the past by the AI community 
but now explored from a database perspective.
The results summarized in Tables~\ref{tab:postulates1}, \ref{tab:postulates2}, and~\ref{tab:summaryRepair} give
indications on the behavior and complexity of 
inconsistency measures for databases,  
helping the reader to figure out which measure 
is more appropriate for specific applications.

Many other interesting issues concerning inconsistency measures in databases
remain unexplored.
We have dealt with denial constraints, 
a common type of integrity constraint which can express 
for instance equality generating dependencies. 
We plan to extend our work to other types of integrity constraints,
and in particular to inclusion dependencies.
Also, we plan to identify tractable cases for the hard measures,
possibly exploiting connections with work done on inconsistent databases
(as shown for $\I_A$), and devise efficient algorithms and index structures 
for evaluating inconsistency measures.
The inconsistency measures we have considered work at the tuple-level,
without distinguishing inconsistency arising from different attributes,
which is another issues we want to address in the future.
Finally, another interesting direction for future work is 
considering databases with null values.

\bibliographystyle{plain}

\bibliography{refs}

\end{document}